\documentclass[aps,prd,onecolumn,groupedaddress,showpacs,nofootinbib,amssymb]{revtex4-2}
\usepackage[dvips]{graphicx}
\usepackage{amssymb}
\usepackage{amsmath}
\usepackage{graphicx,,color}
\usepackage{amsfonts}
\usepackage{bm}
\usepackage{cancel}
\usepackage{comment}
\usepackage{floatflt}
\usepackage{slashed}
\usepackage{appendix}
\usepackage{array}
\usepackage{tabularx}

\newcommand\be{\begin{equation}}
\newcommand\ee{\end{equation}}

\allowdisplaybreaks[4]

\begin{document}

\title{Low-scale Mirror Standard Model Dark Matter and its Detection via Gravitational Waves and the Guitar Nebula}
\author{V.K. Oikonomou$^{1,2}$}
\affiliation{$^{1)}$Department of Physics, Aristotle University of
Thessaloniki, Thessaloniki 54124, Greece} \affiliation{$^{2)}$L.N.
Gumilyov Eurasian National University - Astana, 010008,
Kazakhstan}


\tolerance=5000

\begin{abstract}
What if the dark matter Sector is truly dark, self-interacting and
unreachable by terrestrial experiments? How could we find hints of
such dark sector if it is experimentally unreachable by any
terrestrial experiment? In this work we study a low-scale mirror
Standard Model which can act as a model for dark matter, which
interacts only gravitationally with the Standard Model particles.
The mirror Standard Model sector particles are stable particles
can comprise a measurable part of the dark matter of the Universe.
These mirror Standard Model particles acquire mass through a
low-temperature dark first order phase transition. We examine in
detail this dark phase transition and we indicate how stochastic
gravitational waves can be generated through this transition. Such
a phase transition can generate stochastic gravitational waves
that can be detected by the future gravitational wave experiments.
For the model we use, the produced energy spectrum of the
gravitational waves can be detected by the Square Kilometer Array.
Moreover, we propose a possible way to detect effects of the
particle nature of dark matter, using observational data coming
from the guitar nebula, which can work if dark matter is
collisional, so interacting dark matter. Without specifying a
model for interacting dark matter, thus choosing an agnostic
approach for interacting dark matter, we assume that the guitar
nebula bow shock is generated by the interaction of the high speed
neutron star that passes through the interstellar medium, which is
assumed to be comprised by interacting dark matter and hydrogen.
Our main proposal is that the opening angle of the bow shock can
be directly related to the speed of sound of the dark matter
particles, and a large angle of the bow shock could be a strong
indicator that the interstellar medium is comprised by collisional
dark matter and hydrogen gas. This is motivated by the Bosma
effect which indicates that hydrogen is always in correlation with
dark matter, and hydrogen gas is strongly present in the guitar
nebula.
\end{abstract}


\maketitle

\section{Introduction}

The mysteries of the early Universe are difficult to be revealed
by terrestrial particle accelerations currently, so the focus of
the scientific community is nowadays turned to the sky.
Specifically, the near future stage-4 Cosmic Microwave Background
(CMB) experiments \cite{CMB-S4:2016ple,SimonsObservatory:2019qwx}
and also the current and future gravitational wave (GW)
experiments like the LISA, SKA, BBO, DECIGO, and the Einstein
telescope
\cite{Hild:2010id,Baker:2019nia,Smith:2019wny,Crowder:2005nr,Smith:2016jqs,Seto:2001qf,Kawamura:2020pcg,Bull:2018lat,LISACosmologyWorkingGroup:2022jok},
are highly anticipated to provide information regarding the early
Universe. Already in 2023 a major observational breakthrough was
achieved, since the NANOGrav and other PTA collaborations
\cite{nanograv,Antoniadis:2023ott,Reardon:2023gzh,Xu:2023wog}
verified the existence of a stochastic GW background. This can be
due to astrophysical sources, like supermassive black hole
mergers, or due to cosmological reasons.

However, a long standing problem in modern theoretical physics, is
the dark matter (DM) problem. To date, this elusive major
component of the Universe remains undetected, and its particle
nature is questioned. Alternatives to DM exist, like Modified
Newtonian Dynamics (MOND) theories, but there are many conceptual
problems in these theories, like for example the lack of a
consistent relativistic framework compatible with the GW170817, or
the inability to explain the CMB and large scale structure
formation. However, there exist in the literature several recent
interesting relativistic frameworks that may mimic the MOND
effects
\cite{Deffayet:2024ciu,Boran:2017rdn,Deffayet:2014lba,Deffayet:2011sk},
however these theories must also explain several gaps that DM
fills very well, like the matter perturbation issues, baryon
acoustic oscillations and other minor or major issues related with
the CMB.

Regarding DM, there is a not so well popular class of DM theories,
which have incredibly interesting phenomenological features, the
mirror DM theories firstly introduced in \cite{Kobzarev:1966qya}
and further developed in Refs.
\cite{Hodges:1993yb,Foot:2004pa,Berezhiani:2003wj} see also Refs.
\cite{Silagadze:2008fa,Foot:2000tp,Chacko:2005pe,Berezhiani:2000gw,Blinnikov:2009nn,Tulin:2017ara,Mohapatra:2001sx,
Blinnikov:1982eh,Blinnikov:1983gh,Foot:2016wvj,Foot:2014osa,Foot:2014uba,Silagadze:2008fa,
Blinnikov:2009nn,Tulin:2017ara,Blinnikov:1982eh,Blinnikov:1983gh,Foot:2016wvj,Foot:2014osa,Foot:2014uba,
Foot:2004pq,Foot:2001ft,Foot:2004dh,Foot:1999hm,Foot:2001pv,Foot:2001ne,Foot:2000iu,Pavsic:1974rq,Foot:1993yp,Ignatiev:2000yy,Ignatiev:2003js,
Ciarcelluti:2004ik,Ciarcelluti:2004ip,Ciarcelluti:2010zz,Dvali:2009fw,Foot:2013msa,Foot:2013vna,Cui:2011wk,Foot:2015mqa,Foot:2014mia,Cline:2013zca,Ibe:2019ena,
Foot:2018qpw,Howe:2021neq,Cyr-Racine:2021oal,Armstrong:2023cis,Ritter:2024sqv,Mohapatra:1996yy,Mohapatra:2000qx,Goldman:2013qla,Berezhiani:1995am}
for further applications. In this article we shall assume that a
measurable part of the DM of the Universe is comprised by
low-scale of mirror DM, with ``low-scale'' referring to sub-GeV
scale. In our model, the DM particles can actually be low-scale
stable mirror SM particles. These type of theories are basically
self-interacting DM theories, but with potentially interesting
phenomenological features. Ordinary mirror DM theories can behave
as collisionless DM at very large scales (like galactic
superclusters) and also like collisional at galactic scales, so
problems like the cusp-core problem, the diversity problem of
rotation curves for spiral galaxies and the too-big-to-fail
problems, are nicely explained in the context of ordinary mirror
DM theories. Thus it might be possible that our low-scale mirror
DM model may also have some implications for these large scale
issues we mentioned, although we just mention this prospect and we
will not go in details for this, this could be part of some future
focused work. The mirror SM particles are in thermal equilibrium
but do not interact with the SM particles, and also their
equilibrium temperature is lower than that of the SM particles
\cite{Foot:2004pa}, for phenomenological reasons related with the
nucleosynthesis. One of the assumptions we made is that the
low-scale mirror DM sector has a distinct vacuum compared to the
ordinary Higgs vacuum, and the mirror SM sector acquires mass once
the mirror SM symmetry breaks, at a low temperature, at which a
first order phase transition occurs. We calculate in detail the
thermal effective potential of our low-scale mirror SM and we
prove that the density parameter contribution of the mirror SM
world is $\Omega_{M}=0.10$, hence a considerable part of the total
DM density parameter is described by mirror DM particles. We also
calculate the implications of this dark phase transition on the
energy spectrum of GWs and we proved that such a transition may be
detected by future GW experiments. Overall, this detection depends
on the specific masses and couplings of the model, and in our case
the model can produce a stochastic GW signal which can be detected
by the Square Kilometer Array.

Furthermore, we provide a possible way to detect the particle
nature of DM, using observations coming from the guitar nebula.
Specifically, we assume that the guitar nebula is generated by the
interaction of the high speed neutron star passing through the
interstellar medium which is mainly comprised by collisional
(self-interacting) DM and hydrogen. Our approach is agnostic
regarding the interacting DM model, thus it is model-independent.
The only assumption is that the interstellar medium is comprised
by neutral hydrogen and self-interacting DM, not necessarily
low-scale mirror DM. We show that the opening angle of the bow
shock can be related to the speed of sound of the DM particles,
and a large angle may indicate that indeed the interstellar medium
is comprised by some collisional DM and hydrogen. Several
phenomenological aspects of mirror DM theories and in general
self-interacting DM theories, in addition to future perspectives
of this work are also discussed.

\section{Overview of Small Mass Scale Mirror SM as DM and its Implied Cosmology}

Mirror DM theory is a not so well known candidate for DM, due to
the fact that DM candidates from supersymmetric SM extensions
overwhelmed the literature for quite some time. However, since
supersymmetry is not observed by the LHC, at least to date, mirror
DM serves as a viable DM candidate. Mirror DM satisfies all the
necessary requirements that a viable DM candidate should satisfy,
that is stability of the DM particles, darkness-non interaction
with SM particles and baryons, the similarity of the ordinary
baryon and DM cosmic abundances, large scale structure formation
and asymptotically flat rotation curves for spiral galaxies.
Mirror DM can have a mild interaction with SM particles, but in
this work we confine ourselves to the case that mirror DM cannot
interact at all with the SM particles and ordinary baryons, only
the gravitational interaction between the two sectors is allowed.

Mirror DM in its original form, was motivated by requiring that
the reflection parity is an exact symmetry in nature. Weak
interactions violate parity, so if a mirror SM existed the parity
would be an unbroken symmetry in nature. This parity would indeed
be an exact symmetry of nature if the mirror SM had the same
electroweak vacuum as the SM, however we will assume that the
mirror SM sector has a distinct vacuum from that of the SM. Thus
the parity in our combined SM and mirror SM theory is
spontaneously broken, see for example \cite{Foot:2000tp}. Thus the
theoretical framework of our model has two copies of the SM, one
being the ordinary SM with gauge group $G=SU(3)\times SU(2)\times
U(1)$ and a mirror low scale SM with gauge group
$G_M=SU_M(3)\times SU_M(2)\times U_M(1)$, hence the total gauge
group of the theory is $G\otimes G_M$. There can be a common grand
unified theory (GUT) origin of the above group, for example by
thinking the fact that $SU(4)$ can be broken in $SU(4)\to
SU(2)\times SU(2)$, the common origin of the theory $G\times G_M$
could be a grand unified theory of the form $SU(3)\times
SU_M(3)\times SU(4)\times U(1)\times U_M(1)$, but we shall not
analyze the possible GUT origin of the particle content of our
model. Some similar line of reasoning but for an entirely
different theoretical context, was given in
\cite{Cembranos:2019yio}. In our context, mirror DM is a low-scale
copy of our world, so cosmologically the mirror world can in
principle evolve in the same way as ordinary SM particles, thus
they can participate in the Big Bang Nucleosynthesis (BBN). It can
be debatable though whether atoms can be formed, since this
strongly depends on the binding energy of atoms like hydrogen,
which in the case of a low-scale mirror DM model can be
significantly smaller than in the ordinary hydrogen case. Thus,
the mirror DM model we shall consider, is a self-interacting DM
model, and stable dark particles comprise the DM. Of course it is
not necessary for the total DM to be comprised by mirror DM, in
fact we will prove that the Universe is comprised by the mirror DM
at an extent of $\Omega_{M}=0.10$, thus the rest of DM can be of
other types of DM, like axions or other viable candidates. Now
coming to the mirror DM we propose, the evolution of mirror DM is
similar to the SM particles, but not identical. The major
difference is in the temperature of mirror DM particles. These are
reheated by the inflaton, and self interactions among the mirror
DM particles keep the mirror SM particles in thermal equilibrium,
but their temperature must be smaller than the temperature of the
ordinary SM particles \cite{Foot:2004pa,Berezhiani:2000gw}. Indeed
due to the fact that the mirror SM particle would contribute the
BBN epoch, thus we would have a double of the standard value for
the density parameter. Also the contribution of mirror neutrinos
and mirror photons would lead to an extra neutrino species $\delta
N_{\nu}=6.14$. Taking into account the fact that $\delta
N_{\nu}=6.14 \left(\frac{T'}{T}\right)^4$
\cite{Foot:2004pa,Berezhiani:2000gw}, where $T'$ and $T$ stand for
the mirror and ordinary SM particles equilibrium temperature, then
by demanding that $\delta N_{\nu}<0.4$ \cite{Planck:2018vyg}
yields $T'<0.5\, T$. We shall assume that $T'=0.5\, T$ so let us
calculate the density parameter of the mirror SM particles in our
Universe. The exact relation between the density parameter of the
mirror SM density parameter $\Omega_{MSM}$ and the ordinary SM
density parameter $\Omega_{SM}$ is \cite{Berezhiani:2000gw},
\begin{equation}\label{relicabundance}
\frac{\Omega_{MSM}}{\Omega_{SM}}=x^3\,D^{-K(x)}\, ,
\end{equation}
where $K(x)$ is $K(x)=\frac{1-x^2}{\sqrt{1+x^4}}$ where
$x=\frac{T'}{T}$, and $D$ is the coefficient $\sim {T'}^2$ of the
finite temperature effective potential for the mirror SM
particles. As we show later, an exact calculation for our model
will yield approximately $\Omega_{MSM}\sim 0.10$, thus a large
part of the present day DM is comprised by mirror DM, recall that
$\Omega_{DM}\sim 0.32$, so almost a third of the DM can be mirror
low-scale DM. Note that $\Omega_{MSM}$ and $\Omega_{SM}$ are total
density parameters for the mirror DM and the ordinary SM baryonic
matter. The important feature is that the equilibrium temperature
of mirror DM is actually smaller than the SM temperature. If an
analogy is used, the cup of coffee is our world, the water mimics
the mirror DM and other DM particles, and sugar and coffee can be
the baryonic world. Thus coffee and sugar are built into the
matter structure offered by water, already existing before sugar
and coffee were created in the world.

The important feature of mirror DM is that it is self-interacting,
which can significantly affect the small galactic scales behavior
of DM, but would unaffect the cluster galactic dynamics. We shall
discuss this issue later on in this section, but also in the
section discussing the guitar nebula phenomenology, the
interacting DM will play a crucial role in the analysis, without
referring to a specific model of self-interacting DM. For a
thorough analysis of the cosmological consequences of mirror DM,
see for example \cite{Berezhiani:2000gw}.

Now let us discuss our motivation for using low-scale mirror SM as
DM, which is a form of collisional DM. Collisional, or
self-interacting DM is motivated by the fact that it may explain
some old and new problems of small scale structure observations
(galactic scales) which are in tension with standard collisionless
cold DM predictions, while at the same time leaving intact the
successes of the $\Lambda$CDM model on large (cluster and
supercluster galactic scales). For a recent review on this see
\cite{Tulin:2017ara}. Self-interacting DM offers a successful
explanation for the cusp-core problem (applied for dwarf and low
surface brightness galaxies), the diversity problem of rotation
curves for spiral galaxies, the missing satellites problem and the
too-big-to-fail problem. We need to comment that on galactic
cluster scales and beyond, collisionless DM offers a nice
description of nature, while it fails to provide a solid
explanation for the small galactic scale problems we just
mentioned.

Low scale collisional DM however does not suppress dissipation
effects for mirror particles, thus this would not prevent the
mirror halo DM to collapse to a disk. This feature however is not
necessarily a disadvantage of the theory. In fact if viewed in
conjunction with the Bosma effect \cite{Labini:2023bfn}, it can
lead to some interesting physics which we now briefly discuss. As
it was shown recently in Ref. \cite{Labini:2023bfn}, DM and HI
distributions at large radii in a spiral galaxy are strongly
correlated. Thus there are strong hints that DM is found in the
disk of spiral galaxy, the roots of this suggestion was the Bosma
effect \cite{Bosma:1981zz}. In effect, the distribution of DM
follows that of HI and it is confined in the disk. This kind of
modelling, can provide a nice fit in several spiral galaxies
\cite{Labini:2023bfn} and thus imposes the question whether DM is
actually found in a disk in spiral galaxies and not in halos. This
is quite compatible with the mirror DM paradigm, especially for
the case that the scale of the mirror DM theory is a low scale,
which is the case we shall consider in this work.

Now this DM candidate is not in the form of atoms but in the form
of stable interacting particles, which is a kind of interacting
DM. Now a basic question is if this DM can clump so that structure
formation can start. The answer to this question is that low-scale
mirror DM can clump on large scales, just as interacting DM
particles clump, but on the contrary on small scales cannot clump
to form stars, because no mirror hydrogen or other atoms exist.
Let us further elaborate on this perspective, interacting DM, even
in the form of low-scale mirror DM interacting particles, can
still clump and form structures despite not being able to form
hydrogen-like atoms.  Here is how interacting DM can clump without
forming hydrogen-like structures, firstly at large scales, like
galactic scales, so multiple Kpc distances, the main interaction
driving DM clumping is gravity. Just like standard cold DM,
interacting DM is subject to gravitational forces, which causes it
to fall into gravitational wells and clump on large scales, such
as in galaxies, galaxy clusters, and DM halos. Also in the early
Universe, gravitational collapse leads to the formation of large
structures, with small DM clumps merging into larger ones over
time. This process, called hierarchical structure formation, and
allows DM to clump on cosmic scales, even without forming atoms.
Now regarding self-interacting DM, in models where DM particles
can scatter off each other via non-gravitational forces, like in
the case of low-scale mirror DM, these self-interactions can lead
to energy redistribution within DM halos. These collisions can
alter the density profiles of DM structures, like for example in
self-interacting DM models, where DM can redistribute energy
through scattering, smoothing out the central regions of DM halos
and leading to cored density profiles, rather than the sharp cusps
predicted by cold DM. Although one must check whether low-scale
mirror DM can indeed provide a solution to the cusp-core problem,
we simply mention the analogy with self-interacting DM models.
While self-interactions will not lead to small, bound structures
like hydrogen atoms, they still allow DM to form clumps at the
halo scale, with particles exchanging momentum and energy within
these halos. Cooling is also possible in mirror low-scale DM,
since these particle can emit mirror photons. In conclusion,
low-scale mirror DM can clump through gravity and, in some cases,
through dark-sector forces, even though it cannot form
hydrogen-like atoms. The clumping occurs mainly on large scales,
such as in galaxies and galaxy clusters, rather than small-scale
structures like stars or planets, due to the absence of atoms that
form gaseous interstellar medium.

\section{Small Mass Scale Mirror SM and the Corresponding Low
Temperature Mirror Electroweak Phase Transition}

\begin{figure}
\centering
\includegraphics[width=35pc]{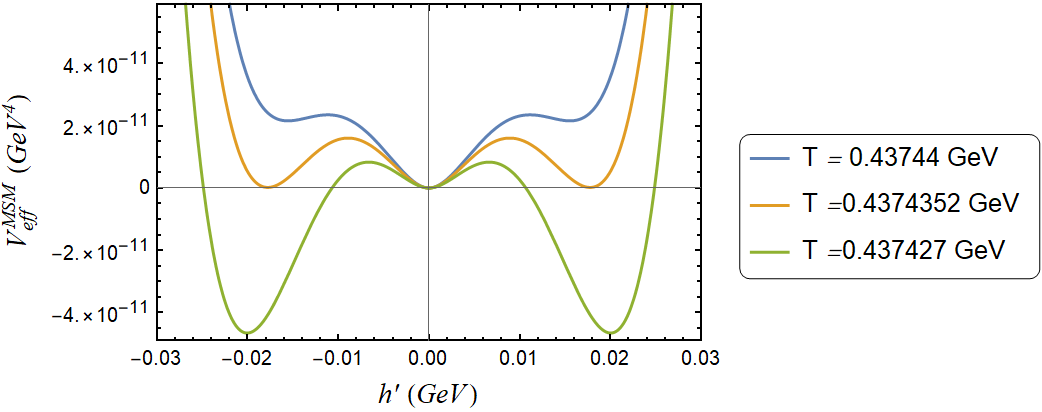}
\caption{The high temperature effective potential of the mirror SM
for various temperatures. The phase transition is a supercooled
phase transition starting at a critical temperature $T^{'}_c\sim
0.4374352\,$GeV and being concluded at an approximate percolation
temperature $T^{'}_*\sim 0.33\,$GeV. }\label{plot1}
\end{figure}

We consider a mirror SM, which contains all the particles of the
SM with a mirror Higgs particle, however with a much smaller scale
of the order $v=0.778\,$GeV. Note that the scale is free to model,
and the value $v=0.778\,$GeV is not of particular importance, it
is  just a toy model choice. We needed a scale of the order
$\mathcal{O}(1)\,$GeV, in order to have a first order phase
transition with the transition temperature being of the order
$\mathcal{O}(1)\,$GeV, so a convenient choice for $v$ is
$v=0.778\,$GeV. When the temperature $T$ in the mirror SM sector
drops below a specific value, a first order phase transition
occurs, and then the particles of the mirror SM sector acquire
masses. We need to stress that the mirror SM is not in thermal
equilibrium with the SM and recall that its temperature is
actually 0.5 times smaller than the SM temperature. The
electroweak phase transition is expected to occur at 100$\,$GeV,
but the mirror SM symmetry breaking is expected to occur at a much
smaller temperature, and for the mirror SM energy scale chosen as
$v=0.778\,$GeV, the temperature for which the transition in the
mirror SM sector occurs is of the order $\sim
\mathcal{O}(1)\,$GeV. Now, let us consider the 1-loop effective
potential of the mirror SM, taking into account the mirror Higgs,
mirror gauge bosons, mirror goldstone bosons and the mirror top
quark, and it has the following form \cite{Anderson:1991zb,
Quiros:1999jp,Arnold:1992rz,Carrington:1991hz,Morrissey:2012db,Dine:1992wr,Dolan:1973qd,Senaha:2020mop},
\begin{equation}\label{eq:A1}
\begin{split}
    V^{MSM}_{eff} (h', T') = & - \frac{\mu^{2}_{H'}}{2} {h'}^2 + \frac{\lambda_{H'}}{4} {h'}^4 + \sum_{i} (-1)^{F_i} n_i \frac{m^4_{i}(h')}{64 \pi^2}\left[ \ln \left( \frac{m^2_{i}(h')}{\mu^2_R}\right) - C_i \right] - \frac{n_t m^4_{t}(h')}{64 \pi^2}\left[ \ln \left( \frac{m^2_{t}(h')}{\mu^2_R}\right) - C_t \right] \\
    & + \sum_{i} \frac{n_iT'^4}{2 \pi^2} J_{B} \left(\frac{m^2_i (h')}{T'^2}\right) - \frac{n_t T'^4}{2 \pi^2} J_{F} \left(\frac{m^2_t (h')}{T'^2}\right) \\
    & + \sum_{i} \frac{\overline{n}_i T'}{12\pi} \left[m^3_i(h') - \left(M^2_i(h',T') \right)^{3/2} \right],
\end{split}
\end{equation}
 where \(i = \{h', \chi', W', Z', \gamma' \}\) corresponds to
the bosons in the mirror SM, and the primes denote that the
particles correspond to the mirror SM sector. Making a
high-temperature expansion of the above effective potential, and
also by including all the daisy graphs, we obtain the following
expression for the 1-loop effective potential of the mirror SM,
\begin{equation}\label{eq:A2}
\begin{split}
    V^{MSM}_{eff} (h', T') = & - \frac{\mu^{2}_{H'}}{2} {h'}^2 + \frac{\lambda_{H'}}{4} {h'}^4+ \frac{m^2_{h'} (h')}{24}{T'}^2 - \frac{T'}{12 \pi} \left[m^2_{h'} (h') + \Pi_{h'} (T')\right]^{3/2} + \frac{m^4_{h'}(h')}{64 \pi^2} \left[\ln \left(\frac{a_b {T'}^2}{\mu^2_R}\right) -\frac{3}{2} \right] \\
    & + \frac{3 m^2_{\chi'} (h')}{24}{T'}^2 - \frac{3T'}{12 \pi} \left[ m^2_{\chi'} (h') + \Pi_{\chi'}(T') \right]^{3/2} + \frac{ 3 m^4_{\chi'} (h')}{64 \pi^2} \left[\ln \left(\frac{a_b {T'}^2}{\mu^2_R}\right) -\frac{3}{2} \right] \\
    & + \frac{6 m^2_{W'} (h')}{24}{T'}^2 - \frac{4T'}{12 \pi} m^3_{W'} (h') - \frac{2T'}{12 \pi}\left[ m^2_{W'} (h') + \Pi_{W_L'} (T') \right]^{3/2} + \frac{ 6 m^4_{W'} (h')}{64 \pi^2} \left[\ln \left(\frac{a_b {T'}^2}{\mu^2_R}\right) -\frac{5}{6} \right] \\
    & + \frac{3 m^2_{Z'} (h')}{24}{T'}^2 - \frac{2T'}{12 \pi} m^3_{Z'} (h') - \frac{T'}{12 \pi} \left[  M^2_{Z_{L}'} (h', T') \right]^{3/2} + \frac{ 3 m^4_{Z'} (h')}{64 \pi^2} \left[\ln \left(\frac{a_b {T'}^2}{\mu^2_R}\right) -\frac{5}{6} \right] \\
    & + \frac{12 m^2_{t'} (h')}{48}{T'}^2 - \frac{ 12 m^4_{t'} (h')}{64 \pi^2} \left[\ln \left(\frac{a_f {T'}^2}{\mu^2_R}\right) -\frac{3}{2} \right] - \frac{T'}{12 \pi}\left[ M^2_{\gamma_{L}'} (h', T') \right]^{3/2},
\end{split}
\end{equation}
where $a_b=223.0993$ and $a_f=13.943$ and with the field-dependent
effective masses being equal to,
\begin{equation}\label{eq:A3}
    m^2_{h'} (h') = - \mu^2_{H'} + 3\lambda_{H'} {h'}^2,
\end{equation}
\begin{equation}
    m^2_{\chi'} (h') = - \mu^2_{H'}+ \lambda_{H'} {h'}^2,
\end{equation}
and also
\begin{equation}\label{effectivemassW}
    m^2_{W'} (h) = \frac{{g'}^2}{4} {h'}^2,
\end{equation}
\begin{equation}\label{effectivemassZ}
    m^2_{Z'} (h) = \frac{{g'}^2 + {\tilde{g}}^{'2}}{4} {h'}^2,
\end{equation}
\begin{equation}\label{effectivemasstop}
    m^2_{t'} (h) = \frac{y^2_{t'}}{2} {h'}^2,
\end{equation}
where \(g'\),\(\tilde{g}^{'}\) and \({y}_{t'}\) are the
\(SU'(2)_L\), \(U'(1)_Y\) and top quark Yukawa couplings,
respectively and $\mu^2_{H'}=0.088$. Furthermore, the
temperature-dependent self-energy corrections for the mirror Higgs
and the mirror Goldstone bosons are equal to,
\begin{equation}\label{eq:A4}
    \Pi_{h'} (T') = \Pi_{\chi'} (T') = \left(\frac{3{g'}^2 }{16} + \frac{{\tilde{g'}}^{2}}{16}  +\frac{y^2_{t'} }{4} + \frac{\lambda_{H'}}{2}\right) {T'}^2
\end{equation}
and in addition, the thermal masses of the mirror gauge bosons are
equal to
\begin{equation}\label{T-W}
    \Pi_{W_L} (T) = \frac{11}{6}{g'}^2 {T'}^2,
\end{equation}
\begin{equation}\label{Z-thermalmass}
    M^2_{Z_L} = \frac{1}{2} \left[ \frac{1}{4} \left({g'}^2 + {\tilde{g'}}^{2}\right) {h'}^2 + \frac{11}{6} \left({g'}^2 +{\tilde{g'}}^{ 2} \right) {T'}^2 + \sqrt{\left({g'}^2 - {\tilde{g'}}^{ 2} \right)^2 \left( \frac{1}{4} {h'}^2 + \frac{11}{6}  {T'}^2 \right)^2  + \frac{{g'}^2 {\tilde{g'}}^{ 2}}{4} {h'}^4 }\right],
\end{equation}
\begin{equation}\label{Photon-thermalmass}
    M^2_{\gamma_L} = \frac{1}{2} \left[ \frac{1}{4} \left({g'}^2 + {\tilde{g'}}^{2}\right) {h'}^2 + \frac{11}{6} \left({g'}^2 + {\tilde{g'}}^{ 2} \right) {T'}^2 - \sqrt{\left({g'}^2 - {\tilde{g'}}^{ 2} \right)^2 \left( \frac{1}{4} {h'}^2 + \frac{11}{6}  {T'}^2 \right)^2  + \frac{{g'}^2 {\tilde{g'}}^{2}}{4} {h'}^4 }\right].
\end{equation}
We shall assume that the mirror Higgs self coupling and the mirror
top quark Yukawa coupling and also the mirror gauge boson
couplings are the following, $\lambda_{H'}=0.01288$,
$y_{t'}=0.314223$ $g'=0.206521$, and \({\tilde{g'}}=0.11058\)
respectively. So the low-scale mirror DM Yukawas are actually
smaller to the ones of the SM which are $y_{t}=0.99366$
$g=0.65307$, and \(g'=0.34968\). Also, with the mirror energy
scale being $v=0.778\,$GeV, the masses of the mirror SM particles
after the mirror SM phase transition occurs, are given in Table
\ref{table:1}.
\begin{table}[h!]
\centering
\begin{tabularx}{0.5\textwidth} {
  | >{\centering\arraybackslash}X
  | >{\centering\arraybackslash}X
   | }
 \hline
Particle & Mass (GeV) \\
 \hline
 $h'$ & $m_{h'}=0.125\,$GeV \\
 \hline
 $W'$ & $m_{W'}=0.0804\,$GeV \\
 \hline
 $Z'$ & $m_{Z'}=0.0912\,$GeV \\
 \hline
 $t'$ & $m_{t'}=0.173\,$GeV \\
 \hline
\end{tabularx}
\caption{Masses of the mirror SM particles, considering only the
mirror top-quark from the fermions.} \label{table:1}
\end{table}
For the phase transition study that follows, we assume that $\mu_R
= 2\,m_{t'}$. Now let us proceed to the study of the phase
transition for the low-scale mirror SM effective potential, and we
shall also investigate whether it is a strong phase transition or
not. In Fig. \ref{plot1} we plot the finite temperature effective
potential of the mirror SM for various temperatures. It is
apparent that the phase transition is first order and the critical
temperature where the two vacua are equivalent is $T^{'}_c\sim
0.4474352\,$GeV. Note that in the plot of Fig. \ref{plot1} we used
a precision of 6 digits below 1. The first order phase transition
picture is not affected by the temperature, and is affected only
by the mirror Yukawa couplings and the mirror Higgs vacuum
expectation value. So the reason for using such high precision is
the fact that the finite temperature potential is very sensitive
to the choice of the temperature near critical point. In order to
pinpoint the critical point we needed high precision determining
the slightest changes in the temperature. But the first order
phase transition is not affected by the precision we used in
pinpointing the critical temperature.

Thus the phase transition is a supercooled phase transition, which
occurs well after the two inequivalent vacua are formed, at an
approximate percolation temperature $T^{'}_*\sim 0.33\,$GeV, and
proceeds by vacuum penetration between the two vacua. The
procedure is standard in the literature and we shall consider only
bubble nucleation effects. In the supercooled phase, the bubbles
of the new phase expand and collide, converting the old phase to
the new phase. Thus the previously massless mirror SM particles,
acquire a non-zero mass after the phase transition. The collisions
of the expanding bubbles of the new phase are a source of
primordial gravitational waves, and the complete prediction of the
theory at hand for the amount of primordial gravitational waves
produced will be given later on in this section. In some sense,
this first order phase transition we describe is a dark phase
transition, and in the literature such phase transitions have also
been studied, in different contexts though, see for example Refs.
\cite{Hall:2019rld,Shelton:2010ta,Dutta:2010va,Servant:2013uwa}.
Let us see whether the above phase transition described by Fig.
\ref{plot1} is a strong phase transition or not. Let us recall the
sphaleron rate criterion,
\begin{equation}\label{sphaleron_rate}
    \frac{\upsilon_c}{T_c} > 0.6 - 1.4.
\end{equation}
In our case, the fraction $\frac{\upsilon_c}{T^{'}_c}$ is of the
order $\frac{\upsilon_c}{T^{'}_c}\sim \mathcal{O}(1.73629)$. Thus
the phase transition is a strong phase transition. However, the
sphaleron criterion is the first step toward characterizing the
transition strong, and thus further analysis, perhaps numerical
too is needed. We perform an analytical approach on this issue in
the next section. We also need to note, that the mirror SM we
presented can be implemented easily with additional scalars which
can make the phase transition even stronger. Note that the scalars
are allowed to have any mass or couplings to the mirror Higgs
(customary approach) or any other particle, thus the
phenomenological model building is quite rich. Now, regarding this
mirror SM sector dark first order phase transition, it definitely
satisfies the Sakharov criteria \cite{Sakharov:1967dj} which
recall are (i) baryon number violation, (ii) \(C\)-\(CP\)
violation, and (iii) departure from thermal equilibrium.  The
above dark phase transition satisfies these criteria since the
phase transition is of first order \cite{Quiros:1999jp,
Trodden:1998ym, Riotto:1998bt, Riotto:1999yt}. Thus we have baryon
creation ahead of the expanding bubble walls where the mirror
\(CP\) and \(C\) asymmetries in the mirror baryonic particle
number densities can be generated by \(CP\)-violating interactions
of the mirror plasma with the expanding bubble wall of the true
mirror vacuum \cite{Morrissey:2012db}. Accordingly, these
asymmetries are diffused into the symmetric phase in the front of
the bubble wall, at which point they are converted to mirror
baryons by mirror sector sphalerons, in a similar way as in the
ordinary SM \cite{Cohen:1993nk}. In the broken mirror phase, the
rate of sphaleron transitions can in principle be strongly
suppressed in order to avoid the washing out the generated
baryons. Hence, in our low-scale mirror SM, the DM particles which
are basically the mirror SM particles, acquire masses at a mirror
temperature $T'\sim 0.33\,$GeV and since the mirror SM temperature
and the SM temperatures are related as $T'=0.5\, T$, it seems that
the mirror particles acquire a mass when $T\sim 0.66\,$GeV. Also
the mirror baryon asymmetry occurs exactly at $T$. Note that both
of these eras are deeply in the radiation domination era at a high
redshift $z\sim 3\times 10^{12}$ regarding the ordinary SM world.

At this point let us also consider an important issue, namely the
prediction of the mirror DM abundance, compared to ordinary
baryons. Having the effective potential available in Eq.
(\ref{eq:A2}), it is easy to write it in the form,
\begin{equation}\label{extraeqn}
V(h',T')\sim \mathcal{D}({T'}^2-T^{'2}_0){h'}^2\, ,
\end{equation}
at leading order to the temperature expansion. The parameter $D$
in our case is equal to,
\begin{equation}\label{parameterD}
\mathcal{D}=\frac{3}{32}{g'}^2+\frac{1}{32}{\tilde{g'}}^2+\frac{1}{8}y_{t'}^2+\frac{\lambda_{H'}}{8}+\frac{\mu_{H'}^2}{v^2}\,
,
\end{equation}
so using the numerical values for the mirror SM particle masses we
have that $\mathcal{D}\sim 0.01994$. The fraction of the energy
densities of the mirror baryons $\Omega_{B'}$ over the ordinary
baryons $\Omega_B$ is equal to,
\begin{equation}\label{mirrorbaryonstoordinary}
\frac{\Omega_{B'}}{\Omega_B}=x^3\,\mathcal{D}^{K(x)}\, ,
\end{equation}
where $x=\frac{T'}{T}=0.5$, so with $\mathcal{D}=0.01994$, we
obtain $\Omega_{B'}= 2.1576\,\Omega_{B}$ and with $\Omega_B\sim
0.05$ in the present Universe, we have $\Omega_{B'}\sim 0.10$.
Thus a large part of the DM of the Universe can be mirror DM, and
recall that $\Omega_{DM}\sim 0.32$ at present day. An important
comment however is in order regarding the fraction of DM in the
Universe, for a mirror Higgs having a low vacuum expectation
value. As it was shown in Ref. \cite{Bansal:2022qbi}, if the
mirror Higgs vacuum expectation value ratio is much below 3, there
can be no more than a percent level of mirror DM abundance. This
mainly comes from the fact that the interacting DM (not forming
atoms) will scatter with dark radiation and in effect the total DM
density will be suppressed. Thus, cosmology can also set stringent
bounds on the abundance of mirror DM.

\begin{figure}
\centering
\includegraphics[width=20pc]{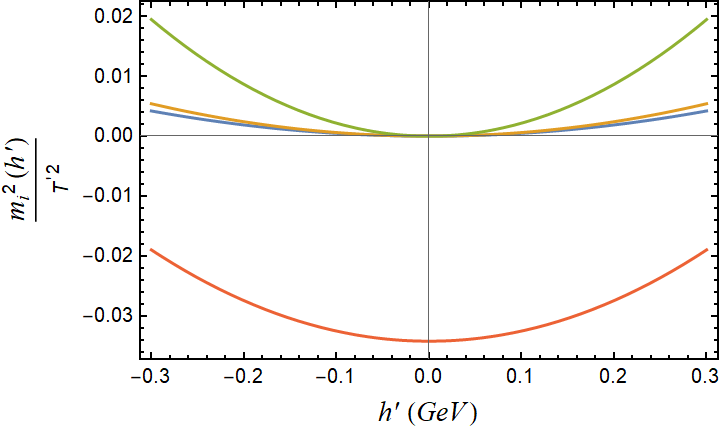}
\includegraphics[width=20pc]{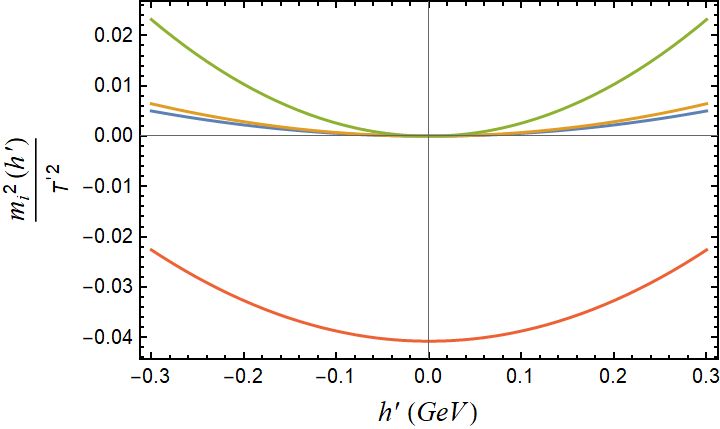}
\includegraphics[width=20pc]{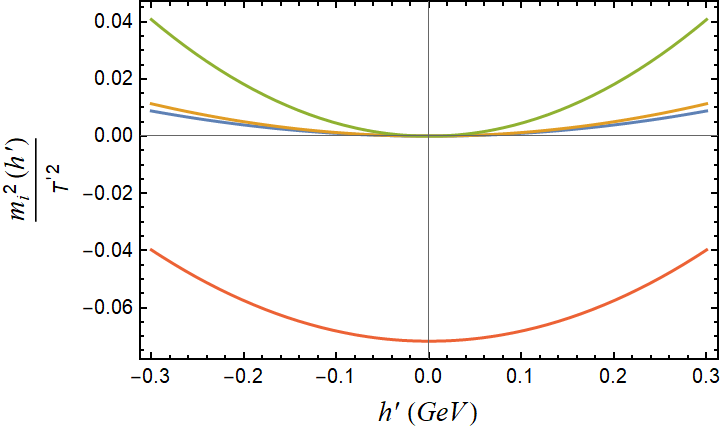}
\caption{The high temperature expansion validation
$m^2_i{h',T'}/{T'}^{2}\ll 1$ for the effective masses of the
mirror W', Z' bosons, the mirror top quark and the mirror Higgs
particle, for $T'=0.4777\,$GeV (upper left) $T'=0.4377\,$GeV
(upper right) and for $T'=0.33\,$GeV (bottom).}\label{plot1a}
\end{figure}

Before closing, let us demonstrate that the high temperature
expansion remains valid for the model we used in this section. In
order for the high temperature expansion to be valid, one must
ensure that the field and temperature dependent masses over the
temperature satisfy $m^2_i(h')/{T'}^{2}\ll 1$. In Fig.
\ref{plot1a} we present the fractions of all the field and
temperature masses over the temperature, and as we can see, the
high temperature expansion approximation remains valid for all the
involved temperatures.

\section{Stochastic GW Signals with the Mirror SM First Order Phase Transition: A Detailed Analysis}

\begin{figure}[h!]
\centering
\includegraphics[width=40pc]{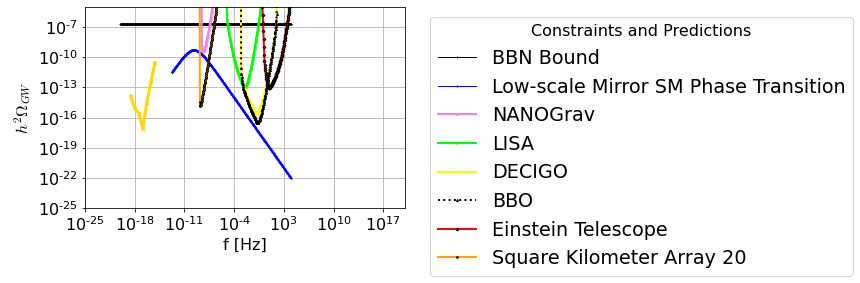}
\caption{The $h^2$-scaled gravitational wave energy spectrum for
the first order phase transition occurring in the low-scale mirror
SM model with effective potential appearing in Eq.
(\ref{eq:A2}).}\label{plot2}
\end{figure}

In the previous section we considered the low-scale mirror SM
phase transition and we found that it starts to occur at a
critical temperature $T_c'\sim 0.4474352\,$GeV so when the
ordinary (baryonic) Universe's temperature is $T\sim 0.894\,$GeV.
Now we shall consider the observational effects of such a phase
transition in terms of stochastic gravitational waves. In the
literature, there exists a vast literature on how stochastic
gravitational waves can be produced in the early Universe
\cite{Kamionkowski:2015yta,Turner:1993vb,Boyle:2005se,Zhang:2005nw,Caprini:2018mtu,Clarke:2020bil,Smith:2005mm,Giovannini:2008tm,Liu:2015psa,Vagnozzi:2020gtf,Giovannini:2023itq,Giovannini:2022eue,Giovannini:2022vha,Giovannini:2020wrx,Giovannini:2019oii,Giovannini:2019ioo,Giovannini:2014vya,Giovannini:2009kg,Kamionkowski:1993fg,Giare:2020vss,Zhao:2006mm,Lasky:2015lej,
Cai:2021uup,Odintsov:2021kup,Lin:2021vwc,Zhang:2021vak,Visinelli:2017bny,Pritchard:2004qp,Khoze:2022nyt,Casalino:2018tcd,Oikonomou:2022xoq,Casalino:2018wnc,ElBourakadi:2022anr,Sturani:2021ucg,Vagnozzi:2022qmc,Arapoglu:2022vbf,Giare:2022wxq,Oikonomou:2021kql,Gerbino:2016sgw,Breitbach:2018ddu,Pi:2019ihn,Khlopov:2023mpo,Odintsov:2022cbm,Benetti:2021uea,Vagnozzi:2020gtf}
including cosmological phase transitions
\cite{Apreda:2001us,Schabinger:2005ei,Kusenko:2006rh,McDonald:1993ex,Chala:2018ari,Davoudiasl:2004be,Baldes:2016rqn,Noble:2007kk,Zhou:2020ojf,
Weir:2017wfa,Hindmarsh:2020hop,Han:2020ekm,Child:2012qg,Fairbairn:2013uta,LISACosmologyWorkingGroup:2022jok,Caprini:2015zlo,Huber:2015znp,
Delaunay:2007wb,Chung:2012vg,Barenboim:2012nh,Senaha:2020mop,Grojean:2006bp,Katz:2014bha,Alves:2018jsw,Athron:2023xlk}.
Let us briefly recall how a first order phase transition can lead
to gravitational wave production and also we shall quote the
energy spectrum of primordial gravitational waves produced by
bubble collisions solely. We are considering a supercooled phase
transition in the thin wall approximation. For the whole phase
transition, there exist bubbles of the old and the new vacuum
state of the Universe. The bubbles of the new vacuum grow with a
bubble wall velocity $v_w$ and we assume that the bubbles of the
new vacuum have a thin wall. These nucleate and grow and thus
collisions between the bubbles of the new vacuum are inevitable.
When the bubbles of the new vacuum grow significantly enough,
these start to collide and coalesce, and the old vacuum is
transformed through this process to the new vacuum. At exactly
this point of bubble collision and coalescence, gravitational
waves are produced. We shall compute the energy spectrum of the
primordial gravitational waves from bubble collisions using the
envelope approximation \cite{Kamionkowski:1993fg} and in the thin
bubble wall limit \cite{Ellis:2020awk}. There are other sources of
stochastic gravitational waves during a first order phase
transition, like sound waves and turbulence, which however we will
not consider in this work. For the energy spectrum of the
primordial gravitational waves during a first order phase
transition we shall use the expression and conventions used by the
NANOGrav collaboration in \cite{NANOGrav:2023hvm}, see also
\cite{Ellis:2020awk,
Xiao:2023dbb,Kamionkowski:1993fg,FitzAxen:2018vdt,Morais:2019fnm}.
Specifically, the energy spectrum of the primordial gravitational
waves caused by the collision of bubbles as a function of the
frequency is,
\begin{equation}\label{bubblecollisionenergyspectrum}
\Omega_b=\frac{\pi^2}{90}\frac{T_0^4}{M_p^2H_0^2}g_*\left(\frac{g_{*,s}^{eq}}{g_{*,s}}\right)^{4/3}\tilde{\Omega}_b\left(\frac{\alpha_*}{1+\alpha_*}\right)^2\left(
H_*R_*\right)^2\,\mathcal{S}(f/f_b)\, ,
\end{equation}
with $\tilde{\Omega}_b=0.0049$, and the spectral function
$S(f/f_b)$ is equal to,
\begin{equation}\label{spectralfunction}
\mathcal{S}(x)=\frac{1}{\mathcal{N}}\frac{(a+b)^c}{(b\,x^{-a/c}+a\,
x^{b/c})^c}\, ,
\end{equation}
where the positive numbers $a$ and $b$ characterize the slope of
the spectrum at the limits of low and high frequencies, and $c$
indicates the width of the peak, while $\mathcal{N}$ is a
normalization constant,
\begin{equation}\label{mathcalN}
\mathcal{N}=\left(\frac{b}{a}
\right)^{a/n}\left(\frac{n\,c}{b}\right)^c\frac{\Gamma(a/n)\Gamma
(b/n)}{n \Gamma (c)}\, ,
\end{equation}
with $n=\frac{a+b}{c}$ and $\Gamma (z)$ is the gamma function. The
values of the slope parameters $a$, $b$ and of the width parameter
$c$ are estimated by numerical calculations, but we will use the
priors used in NANOGrav \cite{NANOGrav:2023hvm} for bubble wall
collisions, so all these parameters take values in the range
$[1,3]$ and specifically $a=1$, $b=1$, $c=3$. Now regarding the
rest of the parameters appearing in the energy spectrum
(\ref{bubblecollisionenergyspectrum}), the parameter $\alpha_*$
characterizes the strength of the phase transition, and values of
the order $\alpha_*\sim \mathcal{O}(0.01)$ indicate a weak
transition, values of the order $\alpha_*\sim \mathcal{O}(0.1)$
indicate an intermediate phase transition, while values of the
order $\alpha_*\sim \mathcal{O}(1)$ or larger, indicate a strong
phase transition \cite{Athron:2023xlk}. The parameters $H_*R_*$
are also important since these are related to the bubble wall
velocity and the duration of the phase transition, since
$H_*R_*=(8 \pi)^{1/3}v_w\,H_*/\beta$, where $\beta$ is a parameter
that characterizes the duration of the phase transition and we
shall calculate this soon, since it is related to the effective
potential, $v_w$ is the velocity of the bubble walls, which we
shall take it equal to unity, following NANOGrav's approach
\cite{NANOGrav:2023hvm}, and $H_*$ is the Hubble radius at the
percolation temperature.  Regarding the bubble wall velocity, it
is an open problem to determine this concretely, so we follow the
modest path and we adopt the value used by the NANOGrav
collaboration corresponding to bubble collisions, as we already
mentioned. Also, for bubble wall collisions, the frequency $f_b$
appearing in Eq. (\ref{bubblecollisionenergyspectrum}) is equal
to,
\begin{equation}\label{peakfrequency}
f_b\simeq 48.5\, nHz\,g_*^{1/2}\left(\frac{g_{*,s}^{e1q}}{g_{*,s}}
\right)^{1/3}\left(\frac{T_*}{\mathrm{GeV}}
\right)\frac{f_b^*R_*}{H_*R_*}\,,
\end{equation}
where $f_{b}^*=0.58/R_*$, where $R_*$ the radius of the wall at
the percolation temperature $T_*$, $g_*$ is the number of
relativistic degrees of freedom, $g_{*,s}$ , and $g_{*,s}^{eq}$
the number of relativistic degrees of freedom contributing to the
entropy at the matter-radiation equality. Now the most important
parameters to be evaluated for the effective potential
(\ref{eq:A2}) are $\alpha_*$ and $\beta$, which recall that
characterize the strength and the duration of the phase
transition. We shall evaluate these in the thin-wall
approximation, using the approach of Ref. \cite{Ellis:2020awk}.
For the calculation of these we need to calculate numerically the
parameter $\sigma$ defined as,
\begin{equation}\label{sigma}
\sigma=\int_{{h'}_{false}}^{h'_{true}}\mathrm{d}h'\sqrt{2V^{SM}_{eff}
(h', 0)-V_{true}}\, ,
\end{equation}
where ${h'}_{true}$ and ${h'}_{false}$ are the values of the
mirror Higgs particle at the false and true vacua respectively.
From this, the Euclidean action $\mathcal{S}_3$ reads,
\begin{equation}\label{euclideanaction}
\mathcal{S}_3=\frac{72\sigma^3T^{'8}}{\alpha_*^2\xi_g^4}\, ,
\end{equation}
where $\xi_g=\sqrt{\frac{30}{\pi^2g_*}}$ and the strength of the
phase transition $\alpha_*$ is evaluated as follows,
\begin{equation}\label{strengthalpha}
\alpha_*=\frac{\Delta V}{\rho_r}\Big{|}_{T=T_*}
\end{equation}
where $\Delta V$ the difference of the true and false vacua in the
effective potential at finite temperature. Now, the duration of
the phase transition is measured by,
\begin{equation}\label{durationphase}
\frac{\beta}{H}=T'\frac{\mathrm{d}}{\mathrm{d}
T'}\left(\frac{S_3}{T'}
\right)=\frac{576\,\sigma^3\,{T'}^8}{\alpha_*^2\xi_g^4}\, .
\end{equation}
For the case of the effective potential (\ref{eq:A2}), and by
assuming that probably the percolation temperature is of the order
$T'_*\sim \mathcal{O}(0.33)\,$GeV, which in the real baryonic
world corresponds to the temperature $T_*\sim 0.66\,$GeV we get
approximately $\alpha_*=0.00106985$ and
$\frac{\beta_*}{H_*}=0.0111423$ so apparently in the thin wall
approximation the transition is relatively weak and fast. In Fig.
\ref{plot2} we plot the energy spectrum of the primordial
gravitational waves (\ref{bubblecollisionenergyspectrum}) as a
function of the frequency. As it can be seen, the predicted signal
can be detected by the Square Kilometer Array. We have to note
though that a detailed numerical evaluation is needed for the
model we used, and also in principle the conventions we used for
the masses and the couplings in the mirror SM sector can be
appropriately chosen so that the gravitational wave signal is
detected by other future experiments. Preliminary results on this
indicate that first order phase transitions from low-scale mirror
DM can be detected only by the Square Kilometer Array, but these
will be presented elsewhere. We do not pursue this analysis
further though in this introductory article and this issue will be
addressed in a future work.
Let us note that low-temperature phase transitions were also
studied in Ref. \cite{Xiao:2023dbb}, but in a different context
and in relation to the NANOGrav 2023 signal. As a final comment,
let us note that in general, in such complicated models, a
detailed numerical analysis is needed and this analysis will
heavily depend on the choices of the couplings and masses of the
mirror SM sector, so we postpone this for a future work, since our
aim in this work was simply to demonstrate the gravitational wave
phenomenology of such low-temperature phase transitions in a
simple and straightforward way. Also it is possible to include a
mirror $SU'(2)$-singlet mirror scalar field with variable mass, in
order to make the phase transition stronger. In these mirror SM
models, there is no restriction on the extra scalar field mass and
couplings to the mirror Higgs particle, thus rich phenomenology
can be obtained. Such an analysis will be performed in a future
work, since this was an introductory work in this kind of
phenomenology.

\section{Finding Direct Imprints of Collisional DM in the Guitar Nebula: An Agnostic and Model Independent Approach}

Now let us proceed to something different in spirit, regarding the
scientific approach. We shall introduce a way which we believe can
provide solid information regarding the particle nature of DM. We
shall consider the possible observational effects that particle DM
can generate to a well-known nebula, the guitar nebula. We shall
not consider a specific model of DM, our main assumption is that
DM is self-interacting, so it can be low-scale mirror SM, like in
the previous section, or some other model. We shall compare the
effects of non-collisional and collisional DM on the guitar nebula
and by this we shall propose a method of identifying indirectly
the presence and effects of DM on the guitar nebula.

The guitar nebula
\cite{cordes,Chatterjee:2003xj,Bucciantini:2018tnu,Gvaramadze:2007de,Tetzlaff:2009uw}
is a spectacular bow-shock in our galaxy, around $1.9$Kpc from the
Sun, created by a rapidly moving neutron star (NS) known as PSR
B2225+65 in the interstellar medium. The speed of the neutron star
is incredible, $v\sim 1600\,$Km$/$sec, thus this supersonic motion
generates the bow shock, which is attributed to the pulsar wind.
In this article we shall make the crucial assumption that the
interstellar medium is a combination of collisional DM and neutral
hydrogen HI. The supersonic motion of the NS creates an accretion
of DM and other interstellar gas components, so we will study the
behavior of the bow shock created by the supersonic motion of the
NS through the DM and hydrogen interstellar medium. As we will
show, the shape of the nebula depends crucially on the average
speed of sound of the interstellar medium, and this can give us
hints on whether one component of the interstellar medium is
actually collisional DM.
\begin{figure}[h!] \centering
\includegraphics[width=25pc]{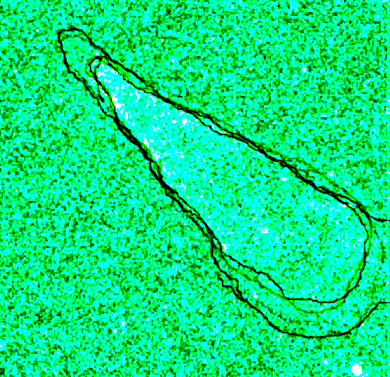}
\caption{Approximate schematic differences of the guitar nebula
from the 1994 and 2001 observations. This is not an actual
astronomic figure, but some graphic representation of the guitar
bow shock.}\label{plot3}
\end{figure}
To start with, in Fig \ref{plot3} we present schematically the
changes in the guitar nebula from its 1994 observation to 2001 and
2006 observations. Fig. 4 is not an astronomic figure, but some
graphic representation of the guitar bow shock. In the following
sections we shall try to explain these changes by using DM
accretion on the NS. The idea is simple, DM and other interstellar
medium gases are accreted on the NS, thus the star feels a
dynamical friction and the matter density of the interstellar
medium around the NS is perturbed and changes with a bow shock
being created. The density behind the NS is maximum, so basically
due to the motion of the rapidly moving neutron star, the bow
shock is filled with DM clumps and hydrogen, which in turn over
the years attract interstellar matter and DM in the bow shock
area. The bow shock looks like an ice cream cone, with the NS
being in its peak, and the density on the sides of the cone and
behind the NS is maximized, so interstellar matter is attracted to
these areas. This way of thinking is motivated by the fact that HI
is a good DM tracker, and since the guitar nebula is observed via
HI, it is tempting to think that DM is actually in the guitar
nebula and plays some important role.

\begin{figure}[h!] \centering
\includegraphics[width=40pc]{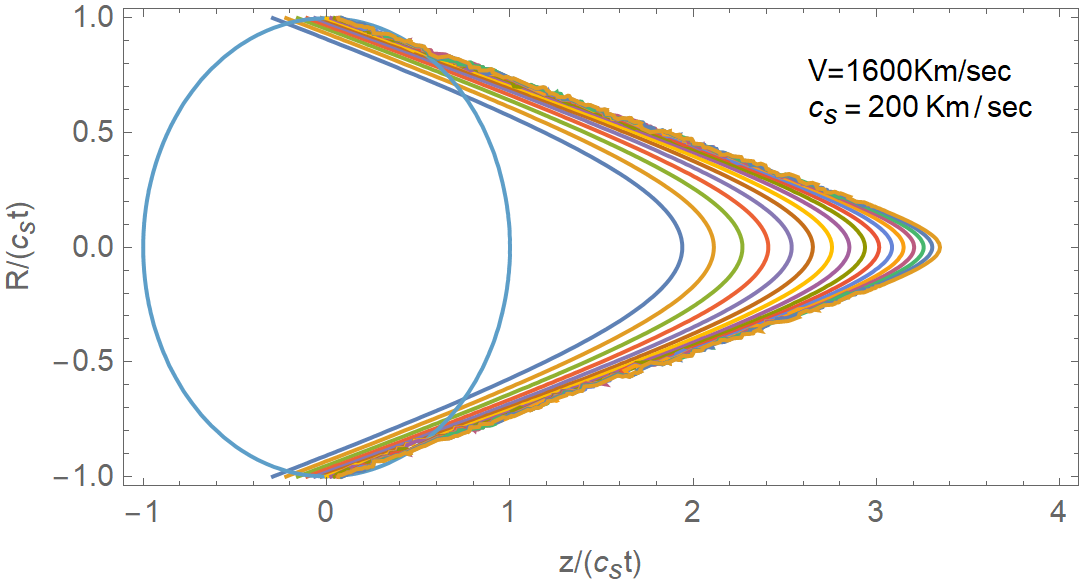}
\includegraphics[width=40pc]{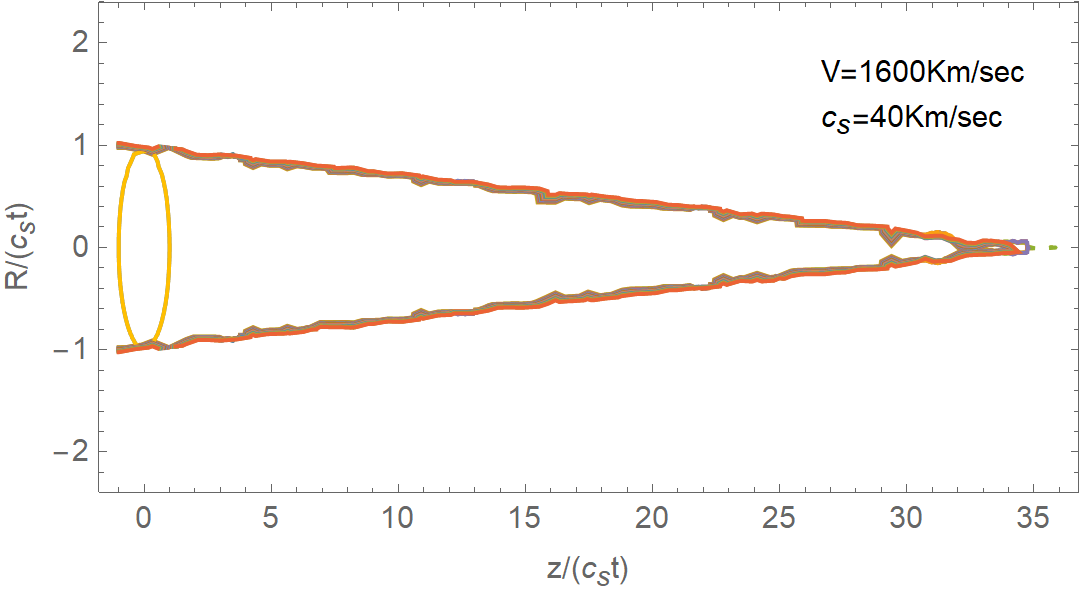}
\caption{The Mach cone of the wake for various values of the sound
speed $c_s$ and for $V=1600\,$Km/sec. The plots correspond to
density perturbation profiles of the supersonic perturber, and
specifically the isosurfaces of $\log \tilde{a}=\log a-\log
(\frac{G M_p}{t c_s^3})$ are presented, with $V=1600\,$Km/sec. The
density increases toward the perturber NS which is situated in the
apex of the Mach cone. The opening angle of the Mach cone $\sin
\theta=\frac{1}{M}=\frac{c_s}{V}$, thus a large opening angle
indicates a large average sound speed for the interstellar medium.
In the upper plot we assumed that the average sound speed is
$c_s=200\,$Km/sec, while in the bottom plot we assumed that the
average sound speed is $c_s=40\,$Km/sec.}\label{plot4}
\end{figure}

Let us discuss how such a scenario can be realized in reality. Let
us consider the linearized equations of motion of a perturbed
density $\rho(x,t)$ of an adiabatic interstellar medium, which in
our case is DM and HI, with average sound speed $c_s$, which is
externally perturbed by the gravitational potential $\Phi(x,t)$,
basically the NSs gravitational potential. If the density is
perturbed by $\rho(x,t)=\rho_0(1+a(x,t))$ and the average medium
sound speed is also perturbed as $u=c_s\beta(x,t)$, then we have
\cite{Ostriker:1998fa},
\begin{equation}\label{main1}
\frac{1}{c_s}\frac{\partial a}{\partial t}+\nabla \beta=0\, ,
\end{equation}
\begin{equation}\label{main2}
\frac{1}{c_s}\frac{\partial \beta}{\partial t}+\nabla
a=-\frac{1}{c_s^2}\nabla \Phi\, ,
\end{equation}
with $c_s=\left( \frac{\partial P}{\partial \rho}\right)^{1/2}$
being the average sound speed of the interstellar medium, $\rho_0$
is the average unperturbed density, and the perturbations satisfy
$a(x,t),\beta (x,t)\ll 1$. The solution for constant speed
perturber with speed $V$ and mass $M$, like the guitar nebula NS,
the density perturbation solution has the following form,
\begin{equation}\label{solutionmain}
a(t)=\frac{G M_p}{c_s^2}\frac{2}{\sqrt{s^2+R^2(1-M^2)}}\, ,
\end{equation}
where we assumed that the movement is along the $\vec{z}$ axis
$\vec{V}=V\vec{z}$, $s=z-V t$, $R=\sqrt{x^2+y^2}$ and
$M=\frac{V}{c_s}$, and $M_p$ is the mass of the perturber. Thus
the perturber is considered like a point mass $M_p$ moving on a
straight line trajectory with velocity $\vec{V}=V\vec{z}$, passing
at $t=0$ through $x=0$. Now it is important to notice that the
supersonic NS perturber generates a density wake only within the
Mach cone which has the following opening angle,
\begin{equation}\label{mainequation}
\sin \theta=\frac{1}{M}=\frac{c_s}{V}\, .
\end{equation}
This is a very important feature of our approach, since a large
opening angle in the Mach cone of the NS perturber will indicate a
large average sound speed of the interstellar gaseous medium and
the opposite, a small sound speed indicates a small opening angle.
Let us proceed with the analysis, so the surfaces of constant
density within the wake correspond to hyperbolae in the $s-R$
plane with eccentricity $e=M$ in the rear of the Mach cone with
$s/R<-(M^2-1)^{1/2}$. Let us present the Mach cone of the wake by
plotting the density perturbation profiles of the supersonic
perturber, and this can be found in Fig. \ref{plot4}, where we
plot the isosurfaces of $\log \tilde{a}=\log a-\log (\frac{G M}{t
c_s^3})$.
As it can  be seen in Fig. \ref{plot4}, the region of the
perturbed density has the shape of a loaded ice cream cone dragged
by its peak point where the perturber NS is. The density increases
toward the perturber NS which is situated in the apex of the Mach
cone, and initially the position of the perturber was at the
origin, in the center of the circles appearing in Fig.
\ref{plot4}. Note that the opening angle of the Mach cone $\sin
\theta=\frac{1}{M}=\frac{c_s}{V}$, determines indirectly the
average sound speed of the interstellar medium. A large opening
angle indicates a large average sound speed for the interstellar
medium. This can be seen clearly in Fig. \ref{plot4}, and
specifically in the upper plot we assumed that the average sound
speed is $c_s=200\,$Km/sec, while in the bottom plot we assumed
that the average sound speed is $c_s=40\,$Km/sec. Clearly the
opening angle in the left plot is significantly larger. This is
our main point and our proposal on how it possible to validate the
presence of particle DM effects on the guitar nebula. Also the
length of the Mach cone in the small sound speed case is quite
larger than the large sound speed case, which is a rather expected
feature, because the NS can travel faster in an interstellar
medium with small sound speed.

Let us elaborate on our argument. The presence of a large sound
speed interstellar medium would be a direct probe of collisional
particle DM. This is simple to understand why, for a temperature
$T\sim 10^4\,$K \cite{Bucciantini:2018tnu}, the hydrogen sound
speed is given by the sound speed of an ideal gas at temperature
$T$ so $c_s=\frac{\gamma k_B T}{m_H}$ where $k_B$ is Boltzmann's
constant, $\gamma$ the adiabatic index and $m_H$ is the hydrogen
mass, so in our case $c_s\sim 11\,$Km/sec. In the case of
collisional DM, we shall adopt the line of research used in Ref.
\cite{Fischer:2024dte}. According to \cite{Fischer:2024dte}, an
estimate of the local sound speed is $v_s\sim 100-300\,$Km/sec,
since we assume that the sound speed is determined only by the
local velocity distribution, which is assumed to obey a
Maxwell-Boltzmann distribution which has an one-dimensional
velocity dispersion $\sigma_{DM}$ \cite{Fischer:2024dte} and it
obeys $c_s=\sqrt{\gamma}\sigma_{DM}$, where $\gamma=5/3$ is the
adiabatic index. For the Milky Way the velocity distribution
$\sigma_{DM}\sim 100-300\,$Km/sec is sufficient. In general, the
speed of sound of DM can be larger than the values we consider in
this work. In general in DM contexts, zero or low values of the
sound speed often reflect the cold nature of DM (low velocity
dispersion and minimal pressure), while high sound speeds indicate
stronger self-interactions. Our approach can indicate if the
interstellar medium is comprised by high or low sound speed
components. A good question is what is the temperature of the DM
particles, however this is very difficult to assess without
referring to a specific model, so we just mention this important
issue.

\begin{figure}[h!] \centering
\includegraphics[width=25pc]{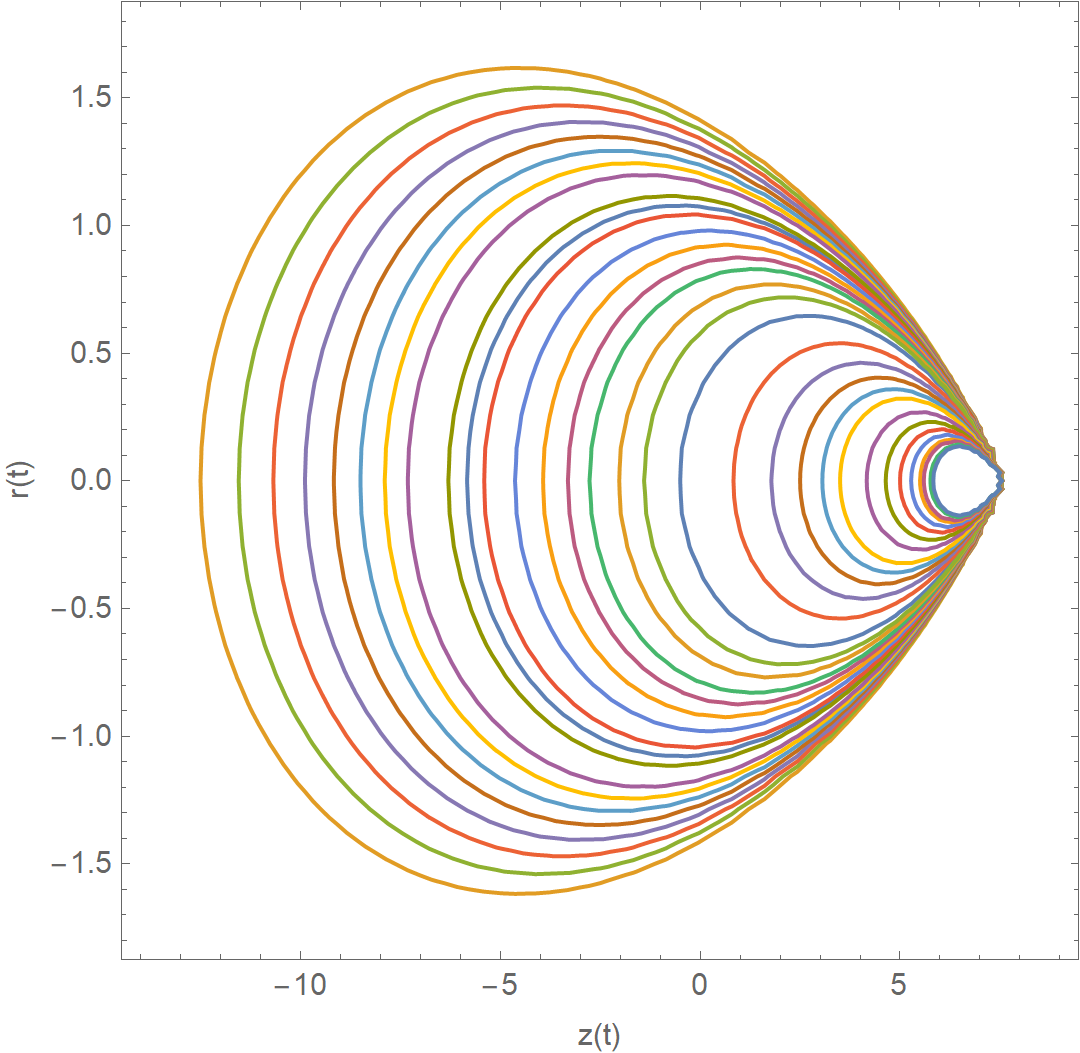}
\caption{The Mach cone of the wake for collisionless DM fluid with
velocity dispersion $\sigma=200\,$Km/sec and for $V=1600\,$Km/sec.
The plots correspond to constant density profiles of the
supersonic perturber.}\label{plot5}
\end{figure}

Let us now consider the collisionless DM case. We compare this
physical picture with the scenario that DM is collisionless and it
accretes on the NS. We shall calculate again the density
perturbations of the collisionless medium, with velocity
dispersion $\sigma$. This calculation can be found in the textbook
\cite{binneytremaine}, and the density response of the wake is,
\begin{equation}\label{collisionless}
\rho(x,t)=\frac{G M_p}{\sigma^2 r}e^{-\frac{V^2\sin^2
\theta}{2\sigma^2}}\left(1-\mathrm{erf}\left(\frac{V}{\cos \theta}
\right) \right)\, ,
\end{equation}
with $r=x-x_M$ and $\theta$ is the angle between $r$ and $V$. The
motion is along the $z$ axis. In Fig. \ref{plot5} we present the
Mach cone of the wake of the collisionless medium, for
collisionless DM fluid with velocity dispersion
$\sigma=200\,$Km/sec and for $V=1600\,$Km/sec. The plots
correspond to density perturbation profiles of the supersonic
perturber. By comparing Figs., \ref{plot4} and \ref{plot5}, the
differences are apparent.
This way of thinking may provide an important hint on whether DM
affects the shape and structure of the guitar nebula. If DM
affects the guitar nebula, and if it is indeed self-interacting,
then Fig. \ref{plot4} is more likely to mimic the guitar nebula,
and it seems schematically that this is true. In fact, if Fig.
\ref{plot4} is an actual description of the reality of the guitar
nebula, then it is not hard to interpret the behavior and
evolution of the guitar nebula. Over the years, the DM and HI
clumps behind the NS, will accrete more DM and HI and thus several
morphological and observational features of the guitar nebula may
be explained. Let us describe one such feature. It was shown in
\cite{Chatterjee:2003xj} that the scale factor of the guitar
nebula changed between the 1994 and the 2001 observations. The
changes in the stand-off radius were interpreted in terms of the
passage of the NS through distinct density areas of the
interstellar medium. We shall give a different interpretation that
does not rely on NS wind physics. The stand-off radius is larger
in the 2001 observation. Assume that DM and hydrogen are accreted
on the NS while it passes through the combined DM-hydrogen medium.
So if DM particles are accreted on the NS, the stand-off radius
is, $R_s=\frac{2 GM_p}{V^2}$, hence the fraction of the 2001 and
1994 observations is,
\begin{equation}\label{r942001}
\frac{R_{1994}}{R_{2001}}=\frac{M_{1994}}{M_{2001}}\left(\frac{v^2_{2001}}{v^2_{1994}}\right)\,
,
\end{equation}
where $M_{1994}$ and $M_{2001}$ are the masses of the NS in 1994
and 2001 respectively, while $v_{1994}$ and $v_{2001}$ are the
speeds of the NS in 1994 and 2001 respectively. Thus since
$R_{2001}>R_{1994}$ this can be explained since
$M_{1994}<M_{2001}$ and $v_{1994}>v_{2001}$. In fact, we have
$\frac{R_{1994}}{R_{2001}}\sim 0.83$ \cite{Chatterjee:2003xj} so
in principle one can perform a detailed calculation. We shall give
some hints on this. The Bondi-Hoyle accretion of DM and HI on the
NS, would increase its mass as follows,
\begin{equation}\label{fe}
\frac{\mathrm{d}M}{\mathrm{d} t}\sim 2\pi
\frac{GM_p^2}{V^{-3}}\rho_0\, ,
\end{equation}
where $\rho_0$ is the average medium density. Now the NS will feel
a drag force on it, which in the case of collisionless DM reads,
\begin{equation}\label{col}
F_{drag}=\frac{4 \pi (G\,M_p)^2}{V^2}\rho_0\left(\frac{1}{2}\ln
\left(1-\frac{1}{M^2}\right)+\ln \left(\frac{V t}{R_{NS}} \right)
\right)\, ,
\end{equation}
assuming moving along with the $z$-axis as in the example we
presented earlier. Note that $R_{NS}$ in Eq. \ref{col} is the
radius of the NS. In principle one can proceed with the details of
this calculations, but it is not our aim to do so, since the drag
force of Eq. (\ref{col}) corresponds to collisionless DM, so it
might not be so accurate, however we wanted to point out that
several morphological aspects of the guitar nebula can be
explained with the accretion of DM and perhaps HI on it. In fact,
the morphological feature in the back of the bow of the nebula may
be explained, since the clumps of DM created in the bow can
attract more DM and neutral hydrogen, and can thus describe this
expansion of the nebula in its rear end and also the brightening
along the bow edges. Notice in Fig. \ref{plot3} the bow edges
which are basically the same and do not expand. In our
perspective, this is due to the fact that the bow density in the
edges is as in Fig. \ref{plot4}. Thus DM and HI is concentrated on
the edges of the nebula and no expansion is predicted in the
edges. The rear however corresponds to places where the NS was in
the past and recall that the density in the back of the NS is
maximum. Thus DM clumps are more dense there, thus the expansion
of the rear end is anticipated. Of course, this is a speculative
approach of ours, and needs to be verified by high-resolution
monitoring observations of the nebula. Our aim was to draw the
scientific community's interest in this speculative perspective.

Let us recapitulate our speculative explanation on the guitar
nebula morphological aspects. The bow shock might be a synergistic
effect of collisional or not DM and neutral hydrogen accretion on
the NS. One can perform exact simulations to determine the sound
speed of the combined DM hydrogen fluid. If DM is collisional then
it is expected to have non-trivial and high sound speed, and also
if the combined effect is given, then powerful simulations taking
into account the hydrogen and DM of collisional nature, might give
an exact explanation of the bow shock nebula form. The magnetic
field wind cannot affect the neutral hydrogen, there might be some
other charged material there, but in the outer skirts of the
galaxy it is more wise to assume that neutral hydrogen and DM
might be there. So one may perform two kinds of simulations, one
with only hydrogen, which has some specific sound speed rather
easy to reproduce at lab, then only DM and then combinations. The
correct simulation will reveal if DM plays some role in the form
of the nebula, how its stand-off radius is decreased and why the
parts in the back increased in such a way between the 1994 and
2004 screenshots of the nebula. I do not have the computing
abilities or scientific knowledge to produce such simulations, so
we leave this work to experts in the specific fields. Our theory
is that hydrogen is attracted to the local DM, which probably is
collisional to some extent, and hydrogen clumps and this might
explain the boundary increase in luminosity in the sides of the
nebula and increase in the size of the rear end of the nebula,
since DM and neutral hydrogen HI is attracted to the local clump
created there. Even combinations of several component interstellar
medium can be simulated to try to produce the shape and evolution
of the nebula during the years 1994 and 2004, this could probably
reveal whether a hidden component of DM with perhaps-if at all-non
trivial sound speed and collisional nature, can participate in the
forming of the guitar nebula and its evolution. If magnetic fields
are not responsible, then a dense unseen medium in conjunction
with the hydrogen composed interstellar medium, might be
responsible for the morphology of the guitar nebula. The density
plots we presented reluctantly indicate that such a physical
picture might be true.

So the point is to reveal the particle nature of DM and if
possible, if it is interacting, like the model we presented in the
previous section, or non-interacting. This can be revealed if the
nebula size and shape can be reproduced with a combination of
gases known to exist in the outer parts of our galaxy. So if the
study reveals a hidden component, this has to be particle DM which
interacts or does not interact. Such a study might actually reveal
the particle nature of DM.

High sound speed velocity of a component of the interstellar
medium at the edge of the galaxy, where the guitar nebula is
situated, could only point to interacting DM. It is hard to find
ordinary matter in the outer galaxy with such high sound speeds.
So simulations will reveal if a dark component is actually there
or not.

\section{Future Perspectives, Discussion on the Results and Conclusions}

In this work we presented a model of DM in which a considerable
portion of the total DM contained in the Universe is comprised by
a low-scale mirror SM. In this model, the DM particles can be
low-mass stable mirror SM particles, thus this type of mirror DM
is self interacting, however in the context of our model mirror
atoms cannot be formed. The low-scale mirror DM particles are in
thermal equilibrium with a smaller temperature compared to the
ordinary SM particles, due to BBN constraints. This type of
collisional, or self-interacting DM is motivated by the fact that
it is possible in this context to explain some old and new
problems of small scale structure observations (galactic scales),
which are currently in tension with standard collisionless cold DM
predictions, and at the same time the same theory leaves intact
the successes of the $\Lambda$CDM model on large (cluster and
supercluster galactic scales). For example, the cusp-core problem,
the diversity problem of rotation curves for spiral galaxies and
the too-big-to-fail problems are nicely explained by collisional
DM models. However, an observational feature of low-scale mirror
DM is that it does not suppress dissipation and thus it is
possible that mirror DM collapses to a disk, like ordinary
baryonic DM. This is a phenomenologically interesting situation in
view of the Bosma effect, which indicates that hydrogen HI is
strongly correlated with DM and it is always found in places where
DM is.

We assumed that the low-scale mirror DM sector has a distinct
vacuum compared to the ordinary Higgs vacuum, thus a
low-temperature first order phase transition can occur in the
mirror DM sector. We calculated the finite temperature effective
potential of this mirror DM sector and we demonstrated that in the
context of our model the total mirror DM density parameter is
$\Omega_{M}=0.10$ thus a considerable part of the total DM density
parameter is described by mirror DM particles. We also studied in
detail the phase transition properties and we demonstrated that
the resulting GW energy spectrum can be detectable by future
Square Kilometer Array.

After that, we discussed how such sort of collisional DM can be
indirectly observed by using observations coming from the guitar
nebula, by studying the morphology of the bow shock. The main
point is that the angle of the bow shock may provide hints for the
sound speed of the interstellar medium, and thus a large sound
speed of the medium can indicate that the interstellar medium
itself is comprised by self-interacting DM. Note that we used an
agnostic and model independent approach for the guitar nebula
study.

Let us discuss some other phenomenological features of low-scale
mirror DM.  An issue is the heating process of the low-scale
mirror SM sector, and a reliable explanation of why low-scale
mirror DM may actually form a disk, like ordinary matter. We
discussed this issue in the previous sections, and we evinced that
actually such a collapse is in principle phenomenologically
allowed and motivated by the Bosma effect. Another good question
is whether low-scale mirror DM received mass before the SM
particles or after. In our case, the mirror DM particles received
mass after the SM particles. In low-scale mirror DM, the low-scale
mirror SM could receive their mass earlier if a high temperature
occurred at a much larger temperature than the electroweak phase
transition one. Also singlet extensions of low-scale or even
high-scale mirror SM models are possible to provide stronger first
order transitions. There also exist many galaxies which are
speculated to be formed solely by DM, that may be well described
by mirror DM, like for example the diffuse galaxy Dragonfly
44-VIRGOHI21 which is inconsistent with the MOND descriptions, and
also FASTJ01139+4328.

Finally, an interesting question is related to the inflationary
era, how is the inflaton coupled to mirror DM, is there a uniform
coupling in the SM and mirror SM, or some distinct coupling, or
even two inflatons. It appears that some non-trivial inflationary
mechanism applies, since the two sectors must have different
temperatures. It is also interesting to see this reheating
phenomenology through the prism of theories with geometric
inflation descriptions, like modified gravity for example.

A challenging phenomenological feature to be explained by
low-scale mirror DM, is related to the Bullet cluster and the
Abell 520 cluster. Ordinary mirror DM can successfully the diverse
galactic mergers like the Bullet cluster and Abell 520. The Bullet
cluster indicates that DM behaves as collisionless DM, while on
the contrary the Abell 520 prefers self-interacting DM. It is
remarkable that ordinary mirror DM can reconcile both these
galactic mergers, since in the case of Abell 520, a large portion
of ordinary mirror DM can be in the form of gas, which is
impossible for WIMPs or other collisionless DM candidates, while
in the case of the Bullet cluster, DM behaves as ordinary stars
\cite{Blinnikov:2009nn}. For a detailed discussion on the diverse
ability of mirror DM to explain galactic scale mergers and
structures, see for example \cite{Silagadze:2008fa}, where the
Bullet cluster, the Abell 520, Hoag's object and other exotic
galactic structures are viewed through the prism of ordinary
mirror DM. Now it is challenging to reconcile these two diverse
observations in the context of low-scale mirror DM, where it is
not possible to have atoms. This is left for future studies
focused on this topic.

The future observations along the lines we described in this work
may reveal the particle nature of the mysterious DM component of
our Universe. For the moment it is very difficult to prove it is
out there directly, in the same way that it is difficult to prove
to someone that the main ingredient of a cup of coffee is not the
brown coffee grinds, that colors coffee and we see it, but water
is the main ingredient.

\section*{Acknowledgments}

This research has been is funded by the Committee of Science of
the Ministry of Education and Science of the Republic of
Kazakhstan (V.K.O) (Grant No. AP19674478).

\end{document}